\documentclass[prl,twocolumn,floatfix,superscriptaddress]{revtex4}
\usepackage{graphicx,amsfonts,amssymb,amsmath,bm}
\usepackage[colorlinks=true]{hyperref}
\usepackage{color}
\usepackage{bm}

\newif\ifhyper
\hypertrue
\ifhyper
\fi

\newlength{\ldag}
\begin{document}

\title{Universal behaviors  in the    wrinkling transition  of    disordered  membranes}

\author{O. Coquand}
\email{coquand@lptmc.jussieu.fr}
\affiliation{Sorbonne Universit\'e, CNRS, Laboratoire de Physique Th\'eorique de la Mati\`ere Condens\'ee, LPTMC, F-75005 Paris, France}
\affiliation{Institut f\"ur Materialphysik im Weltraum, Deutsches Zentrum f\"ur  Luft- und Raumfahrt, Linder H\"ohe, 51147 K\"oln, Germany}

\author{K. Essafi}
\email{essafi@lptmc.jussieu.fr}
\affiliation{Sorbonne Universit\'e, CNRS, Laboratoire de Physique Th\'eorique de la Mati\`ere Condens\'ee, LPTMC, F-75005 Paris, France}

\author{J.-P.  Kownacki}
\email{kownacki@u-cergy.fr}
\affiliation{LPTM, CNRS UMR 8089-Universit\'e de Cergy-Pontoise,   2 avenue Adolphe Chauvin, 95302 Cergy-Pontoise  Cedex, France}

\author{D. Mouhanna}
\email{mouhanna@lptmc.jussieu.fr}
\affiliation{Sorbonne Universit\'e, CNRS, Laboratoire de Physique Th\'eorique de la Mati\`ere Condens\'ee, LPTMC, F-75005 Paris, France}

%------------------------------------------------------------------------------

\begin{abstract}

The wrinkling transition experimentally   identified    by Mutz  {\it et al.} [Phys. Rev. Lett. 67, 923 (1991)]  and  then thoroughly studied  by Chaieb {\it et al.}  [Phys. Rev. Lett. 96, 078101 (2006)]    in partially polymerized lipid membranes is reconsidered. One shows that the features associated  with  this  transition, notably  the various scaling  behaviors  of the height-height correlation functions  that have been observed,   are  qualitatively and quantitatively well described by  a recent nonperturbative  renormalization group  approach to quenched disordered  membranes by Coquand {\it et al.}  [Phys. Rev E 97, 030102(R) (2018)].  As these   behaviors  are associated  with   fixed points of renormalization group  transformations they  are   {\it universal} and   should also be observed in, {\it e.g.},   defective  graphene and graphene-like materials.

\end{abstract}

\maketitle

\section{I. Introduction}

A considerable  activity has been devoted  these last years  to understanding both experimentally and theoretically  the effects  of   disorder in   membranes, mainly  within the contexts  of  the  current  study of  graphene and graphene-like materials  on the one hand  and,  in a more distant past, of
 partially polymerized lipid membranes on the other hand.  Indeed,  the synthesis  of  graphene \cite{novoselov04,novoselov05}  followed by  the discovery of   its  outstanding  mechanical, electronic, optical  and thermal  properties \cite{castroneto09,katsnelson12,amorim16,akinwande17}  has stimulated  intensive researches      aiming  at understanding  how  the unavoidable presence of defects, vacancies,  or adatoms would alter the physical properties  of  pristine   compounds.   Also,  beyond the  mere  presence of  native  imperfections,   the   introduction of artificial defects, {\it e.g.} foreign adatoms or substitutional impurities,   with the help of  various processes  --  particle (electrons or ions)   irradiation,  chemical methods like oxidation  or crystal growth -- has given rise to   the emergence of  a whole  defect engineering industry    aiming at   achieving  new functionalities for these {\it topologically   designed}  graphene  and graphene-like materials \cite{banhart11,liu15,yang18,ni18}.  Among the numerous effects  observed one finds:  variation (increase or decrease) of   electronic conductivity according to the size of the defects,  increase of   elasticity for  moderate  density of vacancies and decrease at higher  density,  decrease of thermal conductance, of  fracture strength, enhancement   of reactivity,  appearance of ferromagnetism and so on    \cite{lopez15,liu15,tian17,yang18,ni18}.
As part of this  defect engineering  activity,   a specific effort  involving various experimental or numerical techniques --  (low pressure) chemical vapor deposition \cite{joo17}, ion/electron  irradiation   \cite{kotakoski11,eder14,pan14,kotakoski15,li17,schleberger18}  or molecular dynamic simulation  \cite{carpenter12,kumar12,ravinder19}  --   has been made toward  the  design of   defect-induced two-dimensional (2D)  amorphous counterparts of graphene and graphene-like materials. A  highlight of this activity is the achievement by  electron irradiation of  a step-by-step, atom-by-atom,  crystal-to-glass  transition   giving rise to a  vacancy-amorphized  graphene  structure \cite{kotakoski11,eder14,joo17}  similar to the   continuous random network proposed by Zachariasen  \cite{zachariasen32}. Many characteristics  of  this transition   have  been determined: the onset of the defect-induced amorphization process, its temperature dependence,  the structural response to vacancy insertion, the nature of the electronic density of states  of the defective configurations   \cite{carpenter12},  a transition  in the fracture response from brittle to ductile when increasing  vacancy concentration  \cite{carpenter13};  finally a careful analysis   of the  glassy-graphene structure  in terms of  a proliferation of nonhexagonal  carbon rings   has been  performed \cite{eder14,ravinder19}.  However we emphasize that   the very  nature of this  glass transition is still unclear. Moreover   there has been,  up to now,  neither within this last  context nor within  the more general  one of the investigation of defective graphene and graphene-like materials, no  characterization of a   quantitative change  between   --  still putative --  ordered and disordered  phases  and   {\it a fortiori} no   indication  of  universal behaviors  associated  with them.

In marked  contrast  with  this situation,  in a very different context, recent investigations of  partially polymerized  lipid membranes by  Chaieb {\it et  al.}\cite{chaieb06},  following the pioneering work of   Mutz {\it et al.}\cite{mutz91},    have led to identify a remarkable  folding-transition while varying  the degree of polymerization.  More precisely  these authors have  shown that,  upon cooling  below  the  chain melting temperature,  partially  polymerized  phospholipid   vesicles  undergo  a  transition from a  relatively  smooth structure, at high polymerization,  to a {\it wrinkled} structure, at low polymerization,  characterized by  randomly frozen normals. This has led  them to suggest that this transition would  be the counterpart of  the spin-glass transition occurring in disordered spin systems  \cite{mutz91,bensimon92}.     Chaieb {\it et al.} \cite{chaieb06}, by considering the height-height correlation functions, have been able  to characterize quantitatively     the various phases  as well as  the  {\it wrinkling} transition  separating  them.   However,  despite the   large amount of theoretical  work  oriented  towards understanding   the physics of disordered membranes,
no  theoretical explanation has been given so far on the grounds of  these results  \cite{nelson91,radzihovsky91b,radzihovsky92,bensimon92,bensimon93,attal93,park96,mori96,benyoussef98,ledoussal93,mori94a,ledoussal18}.

In this article,  we show that a recent  nonperturbative renormaliation group (NPRG) study,  performed by the present authors \cite{coquand18},  of the  effective theory used to study both  curvature and metric   disorders   perfectly accounts for this situation.  In a first part,  we recall  the experimental status of  wrinkled partially polymerized membranes. In a second part,   we lay out    the unusually   unsettled  state of the theoretical  situation. Finally, in a third part,  performing   an analysis of the long-distance   morphology  of membranes at and  in the vicinity of the wrinkling transition,  we show  how the NPRG approach reproduces  the experimental outputs. Finally  we conclude,  stressing   the consequences of our analysis  for the physics of graphene and graphene-like materials and claiming, in particular, that the  behaviors observed in partially polymerized lipid membranes   should also be observed in these materials.

\section{II. Wrinkling transition in partially polymerized  membranes}

\begin{figure}[h]
\includegraphics[scale=0.3]{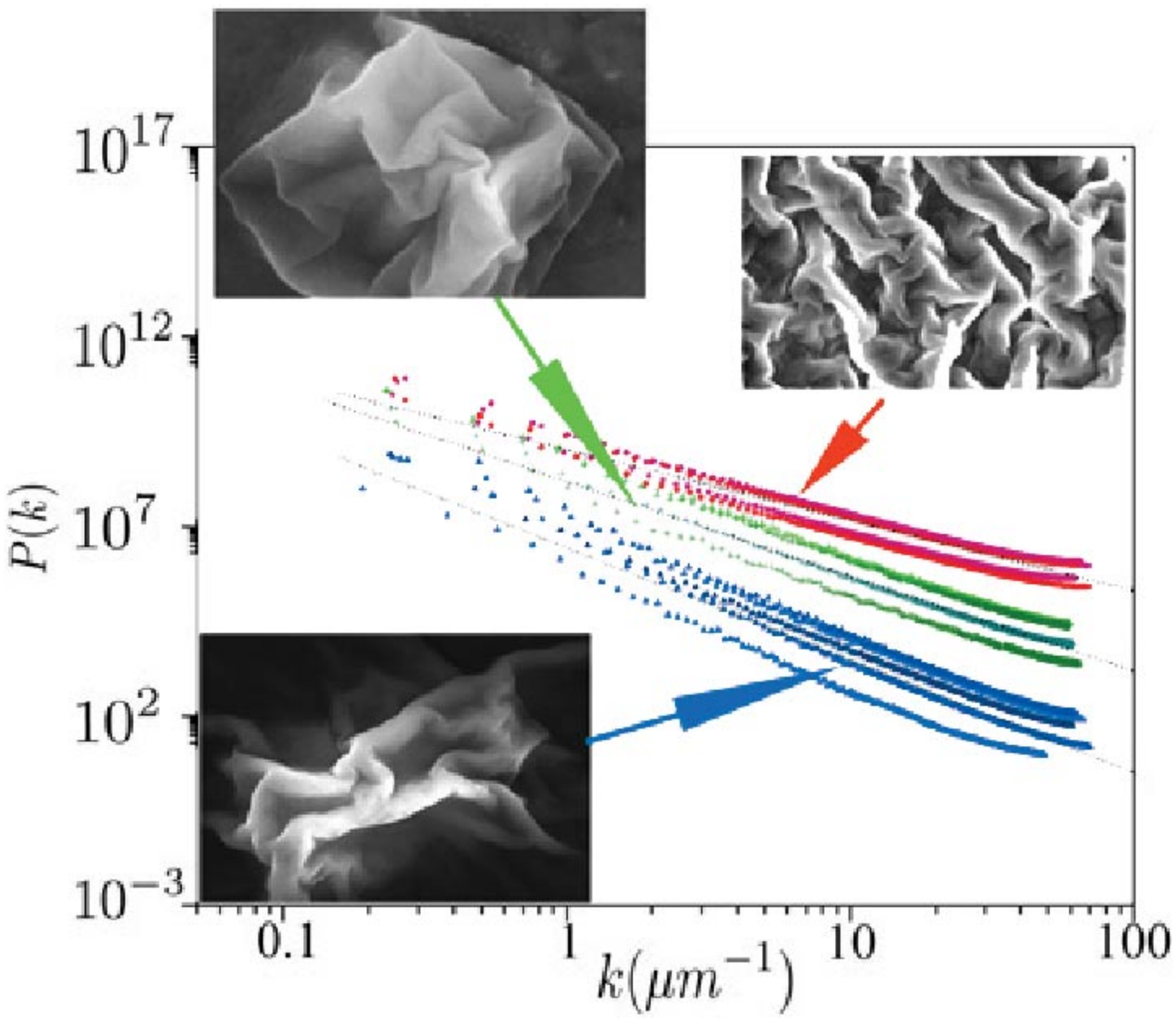}
\caption{The three  scaling behaviors of the power spectrum $P(k)$ as  function of $k$ for various  degrees of polymerization $\phi$ of the membrane. From top to bottom: $\phi=40\%$, $\phi=30\%$ and $\phi=9\%$ and the corresponding membrane configurations.  From Chaieb {\it et al.}   \cite{chaieb06} with permission.}
\label{Fig1}
\end{figure}

The identification   of  a  wrinkling transition in partially polymerized membranes goes  back to the work of  Sackman {\it et al.} \cite{sackman85} on   mixture of diacetylenic phospholipids and dimyristoylphosphatidylcholine,   followed by those  of Mutz  {\it et al.} \cite{mutz91}   and  Chaieb {\it et al.}   \cite{chaieb06}  on  diacetylenic phospholipids       [1,2-bis(10,12-tricosadiynoyl)-{\it sn}-glycero-3-phosphocholine],   who  have  taken advantage of the fact that, upon  a chemical or photochemical process, notably  ultraviolet (UV)  irradiation,  these compounds  polymerize.  In the case considered  in    \cite{mutz91}   the  polymerizable phospholipids are  first prepared  as giant vesicles  and then cooled below the chain melting temperature $T_m\simeq 40^{\circ}$C  where  they form tubular structures that are  then partially polymerized by UV  irradiation.  The membranes are then reheated above $T_m$  where they reform spherical vesicles provided the degree of polymerization does not exceed the percolation threshold located  around 40 $\%$. These vesicles, of typical size ranging from  0.3 to 40 $\mu$m,  are then cooled down to $T_w\simeq 18-22 ^{\circ}$C   where they undergo a spontaneous, reversible,  phase transition  from a  relatively  smooth  structure to a {\it  wrinkled}, highly convoluted,  rigid one displaying  locally  high spontaneous curvature.  This observation has led Mutz  {\it et al.} \cite{mutz91}   to conjecture  that this state of affairs   should be  well described by a theory of polymerized membranes submitted to  quenched curvature disorder.    The outcomes  of  this experiment  have  been made more quantitative by  Chaieb {\it et al.}  \cite{chaieb06, chaieb08,chaieb13,chaieb13bis} who have studied the transition by  various techniques. Small angle neutron scattering has been used to investigate the local structure, giving access to the fractal dimension while environmental scanning electron microscopy  has been  employed   for  the  study of  the  surface topography at  mesoscopic scale  \cite{chaieb06, chaieb08,chaieb13,chaieb13bis}. Finally  a tapping-mode  atomic force microscope   has provided information on the mean-square fluctuations of the surface height  $h({\bf x})$ at a point ${\bf x}=(x_1,x_2)$, relative to the mean surface height, $\langle (h({\bf x})-h({\bf 0}))^2\rangle$,  and  its Fourier transform, the power spectrum  $P(k)$ \cite{chaieb06}. This last quantity  has been    found to  display a remarkable   power-law behavior  in the range $0.1-100$ $\mu$m$^{-1}$: $P(k)\sim k^{-\gamma}$ where the power exponent $\gamma$ \footnote{The notation $\eta$ is employed  in  \cite{chaieb06}  in  place of  $\gamma$ while $\eta$   is used here to indicate the anomalous dimension.}   \  is directly related to   the roughness exponent  $\zeta$ by:  $\gamma=1+2\zeta$.   Three clear distinct regimes   have  been observed  \cite{chaieb06} as the degree of polymerization  $\phi$  is varied,  see Fig.\ref{Fig1}. At low polymerization, typically for  $\phi<30\%$, the surface of the vesicles  presents -- at large scales -- large deformations, creases,  of  order of the vesicle size (500 nm)   typical of a wrinkled state.  In this case one finds \cite{chaieb06}:        $\gamma=2.9\pm 0.1$ corresponding to $\zeta=0.95\pm 0.05$.
 At high polymerization, typically between $32$ and $40\%$  the vesicles  are regular at large scales  and the creases are   less pronounced (of order 20 nm) and   one has    \cite{chaieb06} : $\gamma=2\pm 0.06$ and    $\zeta=0.5\pm 0.03$. Finally,   for $\phi$  in the intermediary region $30\% \le \phi < 32\%$ a  transition occurs and the  vesicles display  the morphology of a crumpled  elastic  sheet with $\gamma=2.51\pm 0.03$ and  $\zeta=0.75\pm 0.02$  \cite{chaieb06}.

%*********************************************************************************************************************************************

%*********************************************************************************************************************************************

\section{III. Theoretical context}

\vspace {0.2cm }

Early investigations of   the wrinkling transition by   Mutz  {\it et al.}   \cite{mutz91}  have  triggered  an impressive series of  theoretical works aiming to  understand   the  effects of  quenched  disorder  contributions  in the seminal  model  of Nelson and Peliti \cite{nelson87}     used   to describe  the flat phase  of  disorder-free polymerized membranes  \cite{proceedings89,bowick01,aronovitz88,guitter89,ledoussal92,kownacki09,gazit09,braghin10,zakharchenko10,roldan11,hasselmann11}.    This series has been initiated by Nelson and Radzihovsky
 \cite{nelson91,radzihovsky91b}  who have mainly investigated  the effects of impurity-induced disorder in the preferred metric tensor. They have in particular shown  that,  below the dimension $D=4$ of the membrane,  the   flat phase of disorder-free membranes  remains stable  at any  finite temperature $T$  but should be  destabilized at vanishing $T$  for any amount of disorder due to a softening of the bending rigidity,  making  possible the emergence  of a spin-glass behavior. This scenario has   been  strengthened by the work of  Radzihovsky and Le Doussal \cite{radzihovsky92}  who, studying  the limit  of large embedding dimension $d$ of the model,  have  identified    an   instability  of the flat phase   toward   a spin-glass-like phase characterized by a nonvanishing Edwards-Anderson order parameter \cite{edwards75}.  At the same time Morse {\it et al.}  \cite{morse92a,morse92b},  extending   the work  done in  \cite{nelson91,radzihovsky91b}, have considered the role of both   curvature and  metric quenched disorders.
Using a perturbative, weak-coupling,   $\epsilon=4-D$ expansion,  they have shown that the  curvature disorder  gives rise  to a new fixed point   at  $T=0$,  {\it stable} with respect to randomness but  {\it unstable} with respect to the temperature.  These works  have been followed by  an intensive search for   various kinds, {\it i.e.}  flat or crumpled, of  glassy   phases  by means of mean-field approximations involving either  short-range  \cite{radzihovsky92,bensimon92,bensimon93,attal93,park96,mori96,benyoussef98} or   long-range    disorders \cite{ledoussal93,mori94a} (see also \cite{ledoussal18} for a review) that have  led   to    predict that,  at sufficiently strong disorder,  the flat phase could  be even unstable  at any  finite temperature   toward a  glassy phase   that would correspond  to the experimentally observed wrinkled phase.  However,  we would like here to emphasize   several  facts. The first one is that these last approaches,  based on  mean-field, large $d$,  computations,  should be considered with great caution when their conclusions are extended  in the finite $d$  case, particularly when they  involve a breakdown of the symmetry between  replica used to perform the average over disorder. Second,  none of the approaches   --  involving    $1/d$   or   $\epsilon=4-D$ expansions  -- has been carried out  at next-to-leading order, where   new physics could emerge.  Finally,   none of them   has  produced   quantitative predictions  or explanations as  for  the results  of Chaieb   {\it et al.} \cite{chaieb06}.

\vspace {0.2cm }

 \section{IV. NPRG analysis}

\vspace {0.2cm }

Recently however, following previous works on  disorder-free polymerized  membranes \cite{kownacki09,braghin10,essafi11,hasselmann11,essafi14,coquand16a} the present authors  \cite{coquand18} have performed a NPRG approach  of  the model considered within a perturbative framework   by Morse {\it et al.}  \cite{morse92a,morse92b} and whose action is given by
\begin{equation}
\begin{array}{ll}
S[\bm{ R}] =\displaystyle \int \text{d}^Dx \hspace{-0.3cm}  & \displaystyle \ \Big\{{\kappa \over 2}\big(\partial^2_{i}\bm{ R(x)}\big)^2
+ {\lambda\over 2}\,  u_{ii}(\bm{ x})^2 +  {\mu}\,  u_{ij}(\bm{ x})^2
\label{action}
\\
&\displaystyle -{\bm c}(\bm{ x}). \partial_i^2\bm{ R(x)}   - \sigma_{ij}(\bm{ x}) \, u_{ij}(\bm{ x})\Big\}\  .
\end{array}
\end{equation}
In this action  $\bm{ R}(\bm{ x})$ is a $d$-dimensional  vector  field  parametrizing, in the  embedding space,   the points $\bm{ x}\equiv x_i$, $i=1\dots D$ of $D$-dimensional membrane while    $u_{ij}$  is the strain  tensor that represents the fluctuations around a flat reference  configuration ${\bf R}^0$: $u_{ij}={1\over 2}(\partial_i \bm{ R}.\partial_j\bm{ R}- \partial_i \bm{ R}^0.\partial_j\bm{ R}^0)$ with \footnote{With a stretching factor (noted $\zeta$  in \cite{coquand18} that should not be confused with the  roughness exponent) taken equal to one.}
\begin{equation}
 \bm{ R}^0=[\langle \bm{ R(\bm{ x})}\rangle]= x_i \bm{ e}_i
\label{groundstate}
 \end{equation}
 where   $\langle\dots\rangle$  and $[\dots]$ denote   thermal and   disorder averages  respectively. In Eq.(\ref{groundstate})  the $\bm{ e}_i$ form  an orthonormal set of $D$ vectors.   The coupling  constants $\kappa$ and $(\lambda,\mu)$ represent respectively the bending rigidity and the Lam\'e coefficients.    The  action (\ref{action})  includes  curvature  and metric   disorder contributions  induced by two random fields ${\bm c}(\bm{ x})$ and $ \sigma_{ij}(\bm{ x})$  that couple to the curvature and strain  tensor respectively. They are considered here as short-range, gaussian fields with  \cite{morse92a,morse92b}
\begin{equation}
\begin{array}{ll}
[c_{i}(\bm{ x})\  c_{j}(\bm{ x}')]= \Delta_{\kappa}\,  \delta _{ij} \ \delta^{(D)}(\bm{ x-x'})
\\
\\
%\end{equation}
%begin{equation}
\big[\sigma_{ij}(\bm{ x})\   \sigma_{kl}(\bm{ x'})\big]= (\Delta_{\lambda} \delta _{ij} \delta_{kl}+2 \Delta_{\mu} I_{ijkl})\ \delta^{(D)}(\bm{ x-x'})
\label{variance}
\end{array}
\end{equation}
where $I_{ijkl}={1\over 2}(\delta_{ik}\delta_{jl}+\delta_{il}\delta_{jk})$, with $i,j,k,l=1\dots D$   where $\Delta_{\kappa}$ and  $(\Delta_{\mu},\Delta_{\lambda}+(2/D) \Delta_{\mu})$ are positive coupling constants associated with  curvature and metric disorders.  Note   finally that the ansatz  (\ref{action}), albeit reduced to  four powers  of the fields  and field derivatives, is expected to lead to  predictions not altered by higher orders, as this happens quite  remarkably   in the disorder-free case \cite{braghin10,hasselmann11,essafi14}    and as this is  strongly suggested  by the very weak sensitivity  of our results with the changes of renormalization group (RG)  process  -- see below.

The RG  equations corresponding to action (\ref{action}) have been  derived  first perturbatively  in \cite{morse92a,morse92b} and then within a NPRG  approach  in \cite{coquand18}. Within this latter approach the RG equations   have  revealed that  there exist,  in   the space of coupling constants,  not only {\it two}, as found by  Morse {\it et al.} \cite{morse92a,morse92b}  but  actually {\it three}  nontrivial fixed points:   the usual   finite-$T$, vanishing-disorder fixed point  $P_4$ associated with  disorder-free membranes \cite{aronovitz88}; a  vanishing-$T$, finite-disorder  fixed point $P_5$    identified for the first time in  \cite{morse92a,morse92b};  finally a finite-$T_c$, finite-disorder  fixed point  $P_c$ found in \cite{coquand18},   missed  within previous   approaches, {\it unstable}  with respect to $T$,    thus  associated with  a  second-order phase transition  and making  the $T=0$ fixed point {\it fully    attractive} provided $T<T_c$.  The consequences of these facts  are twofold: 1)   a whole ``glassy phase''   associated  with  the $T=0$  fixed point   is predicted in agreement with   the wrinkled phase observed in \cite{mutz91,chaieb06}  and 2)   three distinct  universal scaling behaviors are  expected, in agreement with the observations  of Chaieb {\it et al.} \cite{chaieb06}.    The subsequent analysis  shows, moreover,   the  {\it quantitative} agreement between the scaling behaviors predicted and those  observed.    The quantity to consider is the roughness exponent $\zeta$. Let us recall how this quantity is defined in a field-theoretical context.  Let us  decompose ${\bm R}({\bm x})$  around the flat phase configuration $\bm{ R}^0(\bm{ x)}$ according to
%\begin{equation}
${\bm R}({\bm x})=\bm{ R}^0(\bm{ x)}+{\bm u}({\bm x})+{\bm h}({\bm x})$
%\nonumber
%\end{equation}
where ${\bm u}({\bm x})$ 	and ${\bm h}({\bm x})$   are respectively the in-plane --  phonon -- and out-of-plane --  flexuron, -- degrees of freedom  parametrizing  the fluctuations around  $\bm{ R}^0(\bm{ x)}$.   Writing   $\delta {\bm h}({\bm x})={\bm h}({\bm x})-\langle {\bm h}({\bm x}) \rangle$ one  defines  the connected and disconnected correlation functions of the ${\bm h}$ field by:
 \begin{equation}
\big[\langle ({\bm h}({\bm x})-{\bm h}({\bm 0}))^2\rangle\big]=T \chi({\bm x})+ C({\bm x})
\nonumber
\end{equation}
where  $T \chi({\bm x})=[\langle (\delta{\bm h}({\bm x})- \delta{\bm h}({\bm 0}))^2\rangle ]$ and  $C({\bm x})=[\langle {\bm h}({\bm x})-{\bm h}({\bm 0})\rangle^2\big]$  that, as usual,  respectively  measure  the  thermal  and disorder  fluctuations. The long-distance behavior of these correlation functions  is typically given by:  $T \chi({\bm x})\sim \vert {\bm x} \vert^{2\zeta}$ and  $C({\bm x})\sim  \vert {\bm x} \vert^{2\zeta'}$
that define two roughness exponents $\zeta$ and $\zeta'$.  In the same way  correlation functions are  defined for the phonon field with  two roughness exponents $\zeta_u$ and $\zeta'_u$ . They are related   to the previous ones by: $\zeta_u=2\zeta-1$  and  $\zeta'_u=2\zeta'-1$.  Similarly, in  momentum space, writing $\delta {\bm h}({\bm q})={\bm h}({\bm q})-\langle {\bm h}({\bm q}) \rangle$, one defines:
\begin{equation}
 G_{hh}({\bm q})=\big[\langle {\bm h}({\bm q})\bm{h}(-{\bm q})\rangle \big]=T \chi({\bm q})+ C({\bm q})
 \label{corq}
 \end{equation}
where  $T \chi({\bm q})=\big[\langle \delta {\bm h}({\bm q})  \delta \bm{h}(-{\bm q})\rangle \big]$ and  $C({\bm q})=\big[\langle {\bm h}({\bm q})\rangle \langle \bm{h}(-{\bm q})\rangle \big]$
that  behave, at low momenta,  as $\chi({\bm q})\sim q^{-(4-\eta)}$  and   $C({\bm q})\sim  q^{-(4-\eta')}$ where $\eta$ and $\eta'$ are the anomalous dimensions evaluated  at the fixed points of the RG equations.   As a  consequence of  expression (\ref{corq}) the scaling behavior   expected for  the  height-height correlation function   $G_{hh}({\bm q})$ is determined by  the relative value of  $\eta$ and $\eta'$   together with the position of the fixed point:  at finite or at vanishing   $T$ \footnote{See \cite{ledoussal18}  for a careful discussion about the scaling behavior of the  correlation functions.}.
 These exponents  are related to the roughness exponents by:   $\zeta={1\over 2}(4-D-\eta)$ and $\zeta'={1\over 2}(4-D-\eta')$ and to the power exponent  $\gamma=5-D-\eta$ or $\gamma=5-D-\eta'$ depending on the exponent -- $\eta$ or $\eta'$ -- that controls the long-distance behavior.

\begin{figure}[h]
\begin{center}
\advance\leftskip-2.5cm
\advance\rightskip-3cm
\includegraphics[scale=0.35]{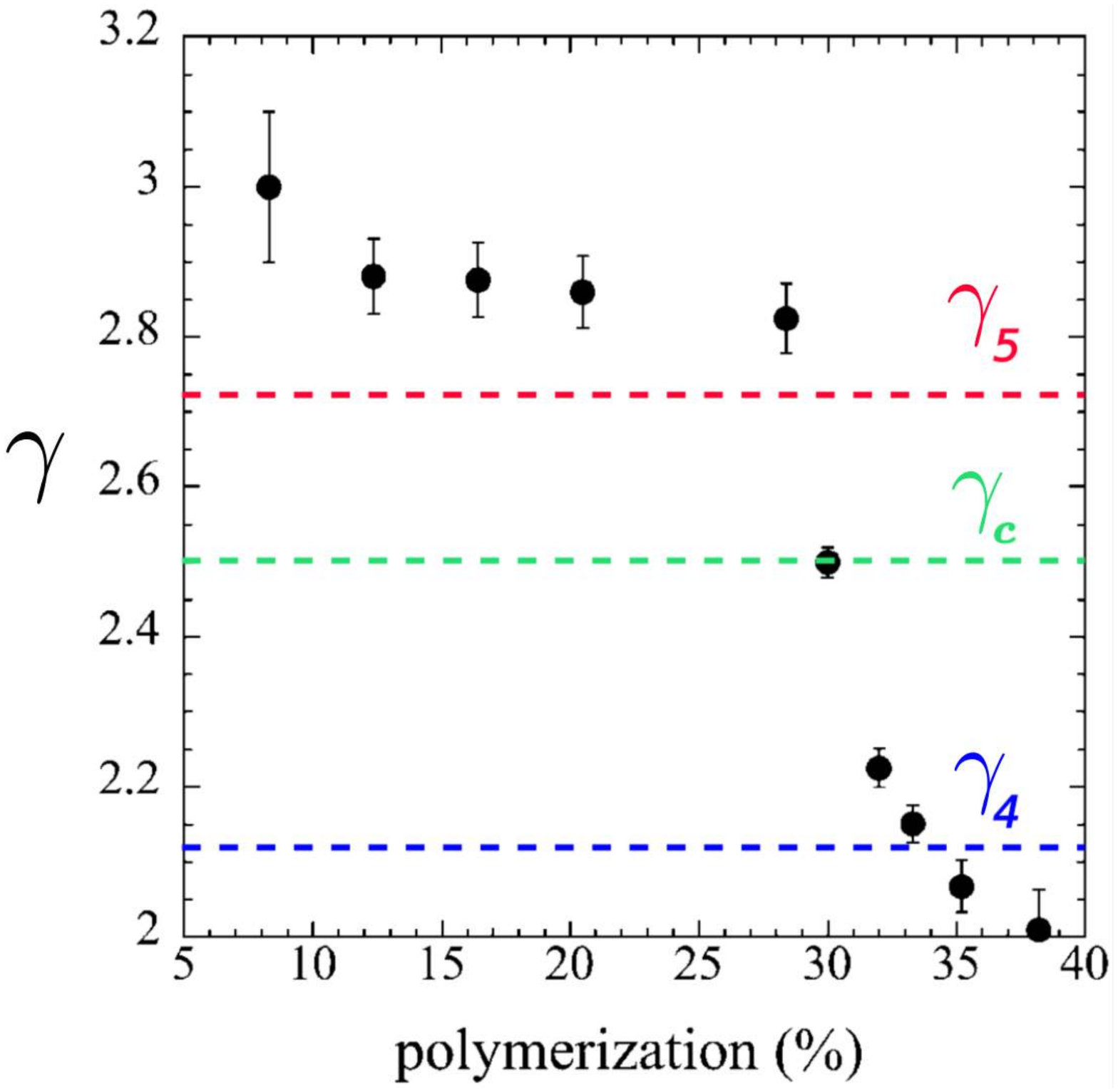}
\caption{Power exponent $\gamma$  as a function of the polymerization rate $\phi$. Adapted from Chaieb {\it et al.}  \cite{chaieb06}. Horizontal lines correspond to the data extracted from  our  NPRG computations: $\gamma_5=2.725(2)$, $\gamma_c=2.510(2)$ and  $\gamma_4=2.151(3)$.}
\label{chaieb}
\end{center}
\end{figure}

At the fully {\it attractive}, vanishing-$T$, finite-disorder   fixed point $P_5$   we  find, by improving the results of \cite{coquand18},  $\eta_5=0.448(2)$ and $\eta_5'=0.275(2)$  \footnote{Error bars  have been  obtained by using  three    families of  cut-off functions  $\widetilde R_k(q)=\alpha \,  Z_k (q^4-k^4) \theta(q^2-k^2)$,     $\widetilde R_k(q)=\alpha\,  Z_k q^4 /  (\hbox{exp}(q^4/k^4)-1)$  and  $\widetilde R_k(q)=\alpha\,  Z_k k^4  \hbox{exp}(-q^4/k^4)$  that are used to separate high and low momenta modes within the RG process (see \cite{coquand18}).  Above,  $\alpha$  is a free parameter used   to investigate the cut-off dependence of  physical quantities and   allows, in particular,   to optimize each cut-off function inside its family, {\it  i.e.}  to  find stationary values of these quantities, see for instance \cite{canet03a}. Error bars follow  from the comparison between the results corresponding to  different optimized  cut-off functions.}. This is in contrast with both  \cite{morse92a,morse92b} and \cite{ledoussal18} where $\eta_5=\eta'_5$ so that $P_5$ was found to correspond   to a {\it marginal}  -- in fact marginally unstable --  fixed point.   As a consequence we  find  a roughness exponent  $\zeta'_5=0.862(1)$   that, according to (\ref{corq})  and the scaling laws of $\chi({\bm q})$ and   $C({\bm q})$, controls the long-distance behavior of the height-height correlation function $G_{hh}({\bm q})$.   This  corresponds  to a power exponent $\gamma_5=2.725(2)$.  This value is very close to that  found in  \cite{chaieb06} at low polymerization -- for $\phi$  in $[10\%;30\%]$ -- and lies  in the range  $[2.80, 2.92]$, see Fig.\ref{chaieb}. As done in \cite{chaieb06}  we exclude the data point at lowest polymerization (corresponding to  $\gamma=3$ in Fig.\ref{chaieb}), which does not belong to the plateau identified for $\phi$  in $[10\%;30\%]$. Note that such a value would correspond to the expected value for a  {\it fluid}  membrane, which could be a hint that below $\phi=10\%$, partially polymerized lipid  membranes do not behave as polymerized membranes anymore. At the  {\it stable},  finite-$T$,  finite-disorder  fixed point $P_4$,    we find  $\eta_4=\eta_4'/2=0.849(3)$ (see  \cite{kownacki09,essafi14,coquand18} and also \cite{amorim16} for a review of other approaches)   that corresponds to $\zeta_4=0.575(2)$ and  $\gamma_4=2.151(3)$.  As seen in Fig.\ref{chaieb} this is again in good agreement with the results found in \cite{chaieb06} at  high polymerization with a value of $\gamma$ that saturates at  2. This value corresponds  to the case of  disorder-free polymerized membranes.  It is  in agreement with  that  obtained by Locatelli  {\it et al.} \cite{locatelli10}   who have found $\zeta=0.54\pm 0.02$ by means of  low energy electron diffraction  on free-standing graphene sheet.  Finally,  at the finite-disorder, finite-$T$,  {\it critical},  fixed point $P_c$ found in  \cite{coquand18}   we  get    $\eta_c=\eta'_c=0.490(2)$ that corresponds to  $\zeta_c=0.755(1)$ and $\gamma_c=2.510(2)$  that is in very good agreement with the  value $\gamma=2.51$ found by Chaieb {\it et al.} \cite{chaieb06} at critical polymerization,  see Fig.\ref{chaieb}.

%*********************************************************************************************************************************************

%*********************************************************************************************************************************************

\section{V. Conclusion}

The conclusion of our work is  fourfold.  First, we have shown that the longstanding problem  of the  wrinkling transition taking place in partially polymerized lipid   membranes is  both qualitatively and quantitatively  clarified  by means of the NPRG approach used in \cite{coquand18}.   Second, reciprocally, this agreement validates  the NPRG approach to disordered polymerized and,  in particular,  consolidates  the prediction    of the existence of three nontrivial fixed points  in the RG flow of the  model (\ref{action}),  in contradiction   with all  previous works. At the formal level this situation   raises the question of the origin of the mismatch between  the NPRG approach and the previous ones: the weak-coupling approach around $D=4$ performed by Morse {\it et al.}  \cite{morse92a,morse92b} and the self-consistent screening approximation    used   by  Radzihovsky and Le Doussal \cite{radzihovsky92,ledoussal18}.   Third,  as the three different kinds  of scaling behaviors  predicted in \cite{coquand18}  are associated  with  fixed points or RG flow, they are universal and should  be observed in a large class of  defective  materials  able to display  curvature disorder \footnote{As observed for the first time by  Morse {\it et al.}  \cite{morse92a,morse92b}, a curvature disorder   generates  metric  disorder.}.   This is in  particular the case of defective graphene, whose sp$^2$-hybridized carbon structure can reorganize into a non-hexagonal structure displaying nonvanishing curvature. Fourth, and finally,    the    glassy graphene  configurations observed during the vacancy-amorphization process  have been shown to display a rough, static, wrinkled structure with reduced thermal fluctuations with  respect to   their purely crystalline counterpart and exhibit  a root mean squared  roughness increasing with vacancy concentration indicating a  change in the  macroscopic morphological/shape structure  of   defective graphene sheets \cite{ravinder19,carpenter12}. It would be of considerable interest  to see if this transition  can be moved closer to the wrinkling transition observed in partially polymerized membranes.

\center{\bf ACKOWLEDGEMENTS}

\begin{widetext}

D.M.    thanks    F. Banhart,  D. Bensimon,  J.-N. Fuchs,  A. Locatelli,  J. Meyer  and O. Pierre-Louis  for fruitful discussions.

\vspace{0.5cm}

\end{widetext}

%\bibliography{bibliotheque1.bib}

\end{document}